# An Analysis of Letter Dynamics in the English Alphabet

Neil Zhao[a,*], Diana Zheng[b]

[a]*Thomas Jefferson University, Philadelphia, Pennsylvania, USA*
[b]*St. Luke's University Health Network, Bethlehem, Pennsylvania, USA*

___________________________________________________________________________________

The frequency with which the letters of the English alphabet appear in writings has been applied to the field of cryptography, the development of keyboard mechanics, and the study of linguistics. We expanded on the statistical analysis of the English alphabet by examining the average frequency which each letter appears in different categories of writings. We evaluated news articles, novels, plays, scientific publications and calculated the frequency of each letter of the alphabet, the information density of each letter, and the overall letter distribution. Furthermore, we developed a metric known as "distance, d" that can be used to algorithmically recognize different categories of writings. The results of our study can be applied to information transmission, large data curation, and linguistics.

_______________________________________________
*E-mail of corresponding author: neil.zhao@students.jefferson.edu

# Introduction

The frequency with which each letter of the English alphabet appears in common writing was estimated by Samuel Morse during the initial development of Morse code [1] and has thereafter found application with different areas of cryptography [2–4]. The tabulation of commonly used letters, as determined by letter frequency, was later utilized to improve typewriter keyboard arrangement by minimizing hand motion [5]. Statistical characteristics of different letters of the English alphabet was further studied in the context of different sentence structures [6]. The letters 'B', 'S', 'M', 'H', 'C' were found to most frequently occur as the initial letters of proper nouns, while 'E', 'A', 'R', 'N' were the most frequently used letters when the entire proper noun is considered. For entire text documents, the most commonly used letters were found to be 'E', 'T', 'A', 'O', 'N'. Interestingly, 95% of the English vocabulary was found to be represented by 13 letters of the alphabet.

Our manuscript expanded upon the statistical study of the English alphabet by evaluating letter frequency in the context of different categories of writings. We analyzed news articles, novels, plays, and scientific articles for letter frequency and distribution. As a result, we determined the information density of the letters of the alphabet. Additionally, we developed a metric called "distance, d" to act as a simple algorithm for recognizing writing category.

## Methods

All analysis was performed with MATLAB R2019a (MathWorks, Natick, MA). Twenty novels and twenty plays were obtained from the online e-books library https://www.gutenberg.org/. Twenty news articles were obtained from https://www.cnn.com/, https://www.foxnews.com/, https://apnews.com/, https://news.yahoo.com/, or https://justthenews.com/. Twenty scientific research articles were obtained from various online journals. Titles of novels, plays, and scientific articles are shown below.

| Novels | Plays | Science |
|---|---|---|
| *Around the World in Eighty Days*, Jules Verne | *Alice in Wonderland*, Alice Gerstenberg | [1] |
| *The Brothers Karamazov*, Fyodor Dostoevsky | *All's Well That Ends Well*, William Shakespeare | [2] |
| *Crime and Punishment*, Fyodor Dostoevsky | *The Importance of Being Earnest*, Oscar Wilde | [3] |
| *The Picture of Dorian Gray*, Oscar Wilde | *The Hairy Ape*, Eugene O'Neill | [4] |
| *Dracula*, Bram Stoker | *The Tragedy of Hamlet*, William Shakespeare | [5] |
| *Emma*, Jane Austen | *The Life of King Henry V*, William Shakespeare | [6] |
| *Ethan Frome*, Edith Wharton | *The Tragedy of Julius Caesar*, William Shakespeare | [7] |
| *Frankenstein; or, the Modern Prometheus*, Mary Shelley | *The Tragedy of King Lear*, William Shakespeare | [8] |
| *Great Expectations*, Charles Dickens | *Macbeth*, William Shakespeare | [9] |
| *The Great Gatsby*, F. Scott Fitzgerald | *The Merchant of Venice*, William Shakespeare | [10] |
| *Grimms' Fairy Tales*, Jacob and Wilhelm Grimm | *The Merry Wives of Windsor*, William Shakespeare | [11] |
| *The Invisible Man*, H. G. Wells | *A Midsummer Night's Dream*, William Shakespeare | [12] |
| *Jane Eyre*, Charlotte Bronte | *Much Ado About Nothing*, William Shakespeare | [13] |
| *Strange Case of Dr. Jekyll and Mr. Hyde*, Robert Louis Stevenson | *Othello, The Moor of Venice*, William Shakespeare | [14] |
| *The Little Prince*, Antoine de Saint-Exupery | *The Life and Death of King Richard the Second*, William Shakespeare | [15] |
| *Pride and Prejudice*, Jane Austen | *The Tragedy of King Richard III*, William Shakespeare | [16] |
| *The Scarlett Letter*, Nathaniel Hawthorne | *The Tragedy of Romeo and Juliet*, William Shakespeare | [17] |
| *Adventures of Sherlock Holmes*, Arthur Conan Doyle | *The Taming of the Shrew*, William Shakespeare | [18] |
| *A Tale of Two Cities*, Charles Dickens | *The Tempest*, William Shakespeare | [19] |
| *The War of the Worlds*, H. G. Wells | *Twelfth Night*, William Shakespeare | [20] |

[1] Lim, F., Solvason, J.J., Ryan, G.E. *et al.* Affinity-optimizing enhancer variants disrupt development. *Nature* (2024). https://doi.org/10.1038/s41586-023-06922-8

[2] Zhao N, Isguven S, Evans R, Schaer TP, Hickok NJ. Berberine disrupts staphylococcal proton motive force to cause potent anti-staphylococcal effects *in vitro*. Biofilm. 2023 Apr 1;5:100117. doi: 10.1016/j.bioflm.2023.100117. PMID: 37090161; PMCID: PMC10113750.

[3] Dastgheyb SS, Hammoud S, Ketonis C, Liu AY, Fitzgerald K, Parvizi J, Purtill J, Ciccotti M, Shapiro IM, Otto M, Hickok NJ. Staphylococcal persistence due to biofilm formation in synovial fluid containing prophylactic cefazolin. Antimicrob Agents Chemother. 2015 Apr;59(4):2122-8. doi: 10.1128/AAC.04579-14. Epub 2015 Jan 26. PMID: 25624333; PMCID: PMC4356782.

[4] Dillon MT, Guevara J, Mohammed K, Patin EC, Smith SA, Dean E, Jones GN, Willis SE, Petrone M, Silva C, Thway K, Bunce C, Roxanis I, Nenclares P, Wilkins A, McLaughlin M, Jayme-Laiche A, Benafif S, Nintos G, Kwatra V, Grove L, Mansfield D, Proszek P, Martin P, Moore L, Swales KE, Banerji U, Saunders MP, Spicer J, Forster MD, Harrington KJ. Durable responses to ATR inhibition with ceralasertib in tumors with genomic defects and high inflammation. J Clin Invest. 2024 Jan 16;134(2):e175369. doi: 10.1172/JCI175369. PMID: 37934611; PMCID: PMC10786692.

[5] Fleming A. On the Antibacterial Action of Cultures of a Penicillium, with Special Reference to their Use in the Isolation of B. influenzæ. Br J Exp Pathol. 1929 Jun;10(3):226–36. PMCID: PMC2048009.

[6] Botting JM, Tachiyama S, Gibson KH, Liu J, Starai VJ, Hoover TR. FlgV forms a flagellar motor ring that is required for optimal motility of Helicobacter pylori. PLoS One. 2023 Nov 17;18(11):e0287514. doi: 10.1371/journal.pone.0287514. PMID: 37976320; PMCID: PMC10655999.

[7] Sun YL, Hennessey EE, Heins H, Yang P, Villacorta-Martin C, Kwan J, Gopalan K, James M, Emili A, Cole FS, Wambach JA, Kotton DN. Human pluripotent stem cell modeling of alveolar type 2 cell dysfunction caused by ABCA3 mutations. J Clin Invest. 2024 Jan 16;134(2):e164274. doi: 10.1172/JCI164274. PMID: 38226623; PMCID: PMC10786693.

[8] Wang X, Yue M, Cheung JPY, Cheung PWH, Fan Y, Wu M, Wang X, Zhao S, Khanshour AM, Rios JJ, Chen Z, Wang X, Tu W, Chan D, Yuan Q, Qin D, Qiu G, Wu Z, Zhang TJ, Ikegawa S, Wu N, Wise CA, Hu Y, Luk KDK, Song YQ, Gao B. Impaired glycine neurotransmission causes adolescent idiopathic scoliosis. J Clin Invest. 2024 Jan 16;134(2):e168783. doi: 10.1172/JCI168783. PMID: 37962965; PMCID: PMC10786698.

[9] Knott S, Curry D, Zhao N, Metgud P, Dastgheyb SS, Purtill C, Harwood M, Chen AF, Schaer TP, Otto M, Hickok NJ. *Staphylococcus aureus* Floating Biofilm Formation and Phenotype in Synovial Fluid Depends on Albumin, Fibrinogen, and Hyaluronic Acid. Front Microbiol. 2021 Apr 29;12:655873. doi: 10.3389/fmicb.2021.655873. PMID: 33995317; PMCID: PMC8117011.

[10] Keenan SW, Beeler SR. Long-term effects of buried vertebrate carcasses on soil biogeochemistry in the Northern Great Plains. PLoS One. 2023 Nov 8;18(11):e0292994. doi: 10.1371/journal.pone.0292994. PMID: 37939031; PMCID: PMC10631625.

[11] de Souza Goncalves L, Chu T, Master R, Chhetri PD, Gao Q, Cil O. $Mg^{2+}$ supplementation treats secretory diarrhea in mice by activating calcium-sensing receptor in intestinal epithelial cells. J Clin Invest. 2024 Jan 16;134(2):e171249. doi: 10.1172/JCI171249. PMID: 37962961; PMCID: PMC10786700.

[12] Zhao N, Curry D, Evans RE, Isguven S, Freeman T, Eisenbrey JR, Forsberg F, Gilbertie JM, Boorman S, Hilliard R, Dastgheyb SS, Machado P, Stanczak M, Harwood M, Chen AF, Parvizi J, Shapiro IM, Hickok NJ, Schaer TP. Microbubble cavitation restores Staphylococcus aureus antibiotic susceptibility in vitro and in a septic arthritis model. Commun Biol. 2023 Apr 17;6(1):425. doi: 10.1038/s42003-023-04752-y. PMID: 37069337; PMCID: PMC10110534.

[13] Sir Isaac Newton's Principia, reprinted for Sir William Thomson LL.D. and Hugh Blackburn M.A. (Glasgow: James Macklehose, publisher to the Univeristy, 1871).

[14] Joo, H.-S. *et al.* Mechanism of Gene Regulation by a Staphylococcus aureus Toxin. *MBio* **7**, (2016).

[15] Evans, C.G., O'Brien, J., Winfree, E. *et al.* Pattern recognition in the nucleation kinetics of non-equilibrium self-assembly. *Nature* **625**, 500–507 (2024). https://doi.org/10.1038/s41586-023-06890-z

[16] Xu J, Zhang Y, Li J, Teper D, Sun X, Jones D, Wang Y, Tao J, Goss EM, Jones JB, Wang N. Phylogenomic analysis of 343 Xanthomonas citri pv. citri strains unravels introduction history and dispersal paths. PLoS Pathog. 2023 Dec 15;19(12):e1011876. doi: 10.1371/journal.ppat.1011876. PMID: 38100539; PMCID: PMC10756548.

[17] Dastgheyb SS, Villaruz AE, Le KY, Tan VY, Duong AC, Chatterjee SS, Cheung GY, Joo HS, Hickok NJ, Otto M. Role of Phenol-Soluble Modulins in Formation of Staphylococcus aureus Biofilms in Synovial Fluid. Infect Immun. 2015 Jul;83(7):2966-75. doi: 10.1128/IAI.00394-15. Epub 2015 May 11. PMID: 25964472; PMCID: PMC4468530.

[18] Wu Q, Badu S, So SY, Treangen TJ, Savidge TC. The pan-microbiome profiling system Taxa4Meta identifies clinical dysbiotic features and classifies diarrheal disease. J Clin Invest. 2024 Jan 16;134(2):e170859. doi: 10.1172/JCI170859. PMID: 37962956; PMCID: PMC10786686.

[19] Zhang W, Sun C, Lang H, Wang J, Li X, Guo J, Zhang Z, Zheng H. Toll receptor ligand Spätzle 4 responses to the highly pathogenic Enterococcus faecalis from Varroa mites in honeybees. PLoS Pathog. 2023 Dec 27;19(12):e1011897. doi: 10.1371/journal.ppat.1011897. PMID: 38150483; PMCID: PMC10775982.

[20] WATSON, J., CRICK, F. Molecular Structure of Nucleic Acids: A Structure for Deoxyribose Nucleic Acid. *Nature* **171**, 737–738 (1953). https://doi.org/10.1038/171737a0

All writings were imported into MATLAB as strings and converted to cells before tabulating the frequency of each letter of the alphabet. All special characters, punctuation, spaces were ignored in the calculation of percentage frequency of letters. Case sensitivity was ignored. Percentage frequency was defined as the ratio of the number of times a letter appeared to the total number of letters, multiplied by 100. The twenty news articles, novels, plays, and scientific articles were stored in a 4 x 21 cell array, with the first column as a header before analysis for percentage letter frequency. The resulting percentages were stored in a 20 x 26 x 4 numerical array. Each element along dimension 1 corresponded to an article of writing; each element along dimension 2 for percentage frequency of each letter of the alphabet; and each element along dimension 3 for each category of writing. MATLAB code provided below with annotation.

```matlab
%fileread to read txt into matlab

function [lettercounter,aveletter,stdletter]=letterfreq(b)
letters='abcdefghijklmnopqrstuvwxyz';
lettercounter=zeros(20,length(letters),4);

for j=1:4
    for k=1:20
        for i=1:length(letters)
            a=num2cell(b{j,k+1});
            lettercounter(k,i,j)=sum(strcmpi(a,letters(i)));
        end
    end
end

%the average frequency of each letter stored in aveletter
%the standard deviation of each average stored in stdletter

aveletter=mean(lettercounter);
stdletter=std(lettercounter);

end

%below is to graph the averages of each letter

for l=1:4
    figure()
    bar(aveletter(1,:,l)/sum(aveletter(1,:,l))*100)
    xticks(1:26)
    ylabel('% of all letters')
    title(names{l,1})
    set(gca,'xticklabel',num2cell(letters))
    hold on
    
    errorbar(1:26,aveletter(1,:,l)/sum(aveletter(1,:,l))*100,stdletter(1,:,l)/sum(avelett
    er(1,:,l))*100,stdletter(1,:,l)/sum(aveletter(1,:,l))*100,'LineStyle','none','Color',
    'k')
    
    hold off
end
```

To calculate the cumulative percentage frequencies of each letter in each writing category, the averages of each percentage letter frequency for all four categories were sorted in descending order before a cumulative summing operation was performed. MATLAB code provided below with annotation.

```matlab
[newssorted,newssortedI]=sort(aveletter(1,:,1)/sum(aveletter(1,:,1))*100,'descend');
[novelsorted,novelsortedI]=sort(aveletter(1,:,2)/sum(aveletter(1,:,2))*100,'descend');
[playsorted,playsortedI]=sort(aveletter(1,:,3)/sum(aveletter(1,:,3))*100,'descend');
[sciencesorted,sciencesortedI]=sort(aveletter(1,:,4)/sum(aveletter(1,:,4))*100,'descend');

figure()
bar(cumsum(newssorted))
xticks(1:26)
ylabel('cumulative % of all letters')
ax1=gca;
set(ax1,'xticklabel',num2cell(letters(newssortedI)))

%to create a second x-axis on top of the graph with different tick markers
ax2=axes('Position',ax1.Position,'XAxisLocation','top','YTick',[],'Color','none');
linkprop([ax1,ax2],{'xlim'})
xticks(ax2,[1:26])
set(ax2,'xticklabel',num2cell(1:26))
ax2.YAxis.Visible='off';

%to elongate the figures
title(names{1,1})
set(ax1,'position',get(ax1,'position').*[1 1 1 0.95])
set(ax2,'position',get(ax2,'position').*[1 1 1 0.95])
set(gcf,'position',get(gcf,'position').*[1 1 1.3 1])

%the above operations are repeated 3 more times for the other cumulative percentage
frequency graphs
figure()
bar(cumsum(novelsorted))
xticks(1:26)
ylim([0 100])
ylabel('cumulative % of all letters')
ax1=gca;
set(ax1,'xticklabel',num2cell(letters(novelsortedI)))

ax2=axes('Position',ax1.Position,'XAxisLocation','top','YTick',[],'Color','none');
linkprop([ax1,ax2],{'xlim'})
xticks(ax2,[1:26])
set(ax2,'xticklabel',num2cell(1:26))
ax2.YAxis.Visible='off';

title(names{2,1})
set(ax1,'position',get(ax1,'position').*[1 1 1 0.95])
set(ax2,'position',get(ax2,'position').*[1 1 1 0.95])
set(gcf,'position',get(gcf,'position').*[1 1 1.3 1])

figure()
bar(cumsum(playsorted))
xticks(1:26)
ylim([0 100])
ylabel('cumulative % of all letters')
ax1=gca;
set(ax1,'xticklabel',num2cell(letters(playsortedI)))

ax2=axes('Position',ax1.Position,'XAxisLocation','top','YTick',[],'Color','none');
linkprop([ax1,ax2],{'xlim'})
xticks(ax2,[1:26])
set(ax2,'xticklabel',num2cell(1:26))
```

```
ax2.YAxis.Visible='off';

title(names{3,1})
set(ax1,'position',get(ax1,'position').*[1 1 1 0.95])
set(ax2,'position',get(ax2,'position').*[1 1 1 0.95])
set(gcf,'position',get(gcf,'position').*[1 1 1.3 1])

figure()
bar(cumsum(sciencesorted))
xticks(1:26)
ylim([0 100])
ylabel('cumulative % of all letters')
ax1=gca;
set(ax1,'xticklabel',num2cell(letters(sciencesortedI)))

ax2=axes('Position',ax1.Position,'XAxisLocation','top','YTick',[],'Color','none');
linkprop([ax1,ax2],{'xlim'})
xticks(ax2,[1:26])
set(ax2,'xticklabel',num2cell(1:26))
ax2.YAxis.Visible='off';

title(names{4,1})
set(ax1,'position',get(ax1,'position').*[1 1 1 0.95])
set(ax2,'position',get(ax2,'position').*[1 1 1 0.95])
set(gcf,'position',get(gcf,'position').*[1 1 1.3 1])
```

The test news, novels, plays, and science articles for evaluating the predictive ability of comparing percentage letter frequencies were obtained from the same sources as above. Titles of the test novels, plays, and scientific articles are shown below.

| **Novels** | **Plays** | **Science** |
|---|---|---|
| *The Call of Cthulhu*, H. P. Lovecraft | *An Ideal Husband*, Oscar Wilde | [1] |
| *The Fall of the House of Usher*, Edgar Allan Poe | *Riders to the Sea*, J. M. Synge | [2] |
| *Robinson Crusoe*, Daniel Defoe | *A Doll's House*, Henrik Ibsen | [3] |
| *The Sun Also Rises*, Ernest Hemingway | *The Game of Chess*, Kenneth Sawyer Goodman | [4] |
| *Ulysses*, James Joyce | *Miss Civilization*, Richard Harding Davis | [5] |

[1] Yadav J, Qadri A. Induction and sustenance of antibacterial activities distinguishes response of mice to Salmonella Typhi from response to Salmonella Typhimurium. Pathog Dis. 2023 Jan 17;81:ftad002. doi: 10.1093/femspd/ftad002. PMID: 36702520.
[2] Valeva SV, Degabriel M, Michal F, Gay G, Rohde JR, Randow F, Lagrange B, Henry T. Comparative study of GBP recruitment on two cytosol-dwelling pathogens, Francisella novicida and Shigella flexneri highlights differences in GBP repertoire and in GBP1 motif requirements. Pathog Dis. 2023 Jan 17;81:ftad005. doi: 10.1093/femspd/ftad005. PMID: 37012222.
[3] Delima GK, Ganti K, Holmes KE, Shartouny JR, Lowen AC. Influenza A virus coinfection dynamics are shaped by distinct virus-virus interactions within and between cells. PLoS Pathog. 2023 Mar 2;19(3):e1010978. doi: 10.1371/journal.ppat.1010978. PMID: 36862762; PMCID: PMC10013887.
[4] Venkatesan A, Jimenez Castro PD, Morosetti A, Horvath H, Chen R, Redman E, Dunn K, Collins JB, Fraser JS, Andersen EC, Kaplan RM, Gilleard JS. Molecular evidence of widespread benzimidazole drug resistance in Ancylostoma caninum from domestic dogs throughout the USA and discovery of a novel β-tubulin benzimidazole resistance mutation. PLoS Pathog. 2023 Mar 2;19(3):e1011146. doi: 10.1371/journal.ppat.1011146. PMID: 36862759; PMCID: PMC10013918.
[5] Klug D, Gautier A, Calvo E, Marois E, Blandin SA. The salivary protein Saglin facilitates efficient midgut colonization of Anopheles mosquitoes by malaria parasites. PLoS Pathog. 2023 Mar 2;19(3):e1010538. doi: 10.1371/journal.ppat.1010538. PMID: 36862755; PMCID: PMC10013899.

All writings were imported as before into MATLAB as strings and converted to cells before tabulating letter frequency. The five news articles, novels, plays, and scientific articles were stored in a 4 x 6 cell array, with the first column as a header before analysis for percentage letter frequency. The resulting percentages were stored in a 5 x 26 x 4 numerical array. Dimension 1 corresponded to an article of writing; dimension 2 for percentage frequency of each letter of the alphabet; and dimension 3 for each category of writing. The metric used for comparing the test articles to the already calculated averages of each letter in each of the four categories of writings was distance. Distance, d, was calculated as follows.

$$d = \sqrt{\sum_i (f_{std,i} - f_{test,i})^2}$$

| $d$ | Distance |
|---|---|
| $f_{std,i}$ | Average percentage letter frequency of standard, iterated through all letters of the alphabet |
| $f_{test,i}$ | Percentage letter frequency of test writing, iterated through all letters of the alphabet |

Distance was interpreted as dissimilarity with the standard under comparison. The greater the distance, the more dissimilar the test writing with the standard. The lower the distance, the more similar the test writing with the standard. MATLAB code provided below with annotation.

```
function [testlettercounter,sqdiff]=letterfreqtest(b,aveletter)

letters='abcdefghijklmnopqrstuvwxyz';
testlettercounter=zeros(5,length(letters),4);
sqdiff=zeros(4,5,4);

for j=1:4
    for k=1:5
        for i=1:length(letters)
            a=num2cell(b{j,k+1});
            testlettercounter(k,i,j)=sum(strcmpi(a,letters(i)));
        end
        %for calculating distance, sqdiff
        diff=testlettercounter(k,:,j)/sum(testlettercounter(k,:,j))*100 
        aveletter(1,:,:)./sum(aveletter(1,:,:))*100;
        sumdiffsq=sqrt(sum(diff.^2,2));
        sqdiff(:,k,j)=reshape(sumdiffsq,4,1);
    end
end
end
```

```matlab
%for graphing the distances
avetest=mean(sqdiff,2);
stdtest=std(sqdiff,0,2);

for i=1:4
    figure()
    bar(avetest(:,1,i))
    ylabel('distance from average')
    title(testnames{i,1})
    set(gca,'xticklabel',testnames(:,1))
    xlabel('comparison standards')
    ylim([0 6]);
    yticks(0:0.5:6);

    hold on
errorbar(1:4,avetest(:,1,i),stdtest(:,1,i),stdtest(:,1,i),'LineStyle','none','Color','k')
    hold off
end
```

The least commonly used 10%, 30%, 50% of letters was determined from the cumulative percentage frequency calculations. Passages of the article *I, Pencil: My Family Tree*, by Leonard E. Read (Read, Leonard. "I, Pencil: My Family Tree" as told to Leonard E. Read, Dec. 1958. Foundation for Economic Education, 1958) were imported into MATLAB as strings. All spaces were replaced with '&'. The letters for 10%, 30%, 50% least commonly used were removed. And all '&' were replaced with '/'. Corresponding passages were also processed to have 10%, 30%, 50% of letters randomly removed. MATLAB code provided below with annotation.

```matlab
red50='hsrdlumwcyfgpbvkxjqzHSRDLUMWCYFGPBVKXJQZ'; %least common 50%
red50=num2cell(red50);
red70='dlumwcyfgpbvkxjqzDLUMWCYFGPBVKXJQZ'; %least common 70%
red70=num2cell(red70);
red90='fgpbvkxjqzFGPBVKXJQZ'; %least common 10%
red90=num2cell(red90);

pass1=replace(pass1,' ','&'); %replace spaces with &
pass1=replace(pass1,red50,' '); %removing least common 10%
pass1=replace(pass1,'&','/'); %replace & with /
writecell({pass1},'pass1.txt') %export passage

pass2=replace(pass2,' ','&');
pass2=replace(pass2,red70,' ');
pass2=replace(pass2,'&','/');
writecell({pass2},'pass2.txt')

pass3=replace(pass3,' ','&');
pass3=replace(pass3,red90,' ');
pass3=replace(pass3,'&','/');
writecell({pass3},'pass3.txt')

%randomly replace 10%,30%,50% of letters
letterscell=num2cell(letters);
letterlocations1=ismember(num2cell(pass1),letterscell); %locating all letters of alphabet
letterlocations2=ismember(num2cell(pass2),letterscell);
```

```
letterlocations3=ismember(num2cell(pass3),letterscell);

length1=length(letterlocations1); %finding size of entire passage
length2=1:length1;
length3=length2(letterlocations1); %find index of each letter
locations1=randperm(length(length3),ceil(length(length3)*0.5)); %locate 50% of letters
locations1=length3(locations1);
pass1=replace(pass1,' ','/'); %replace spaces with /
pass1(locations1)=' '; %replace 50% of letters with spaces
writecell({pass1},'rand50 pass1.txt') %export passage

length11=length(letterlocations2);
length22=1:length11;
length33=length22(letterlocations2);
locations2=randperm(length(length33),ceil(length(length33)*0.3));
locations2=length33(locations2);
pass2=replace(pass2,' ','/');
pass2(locations2)=' ';
writecell({pass2},'rand70 pass2.txt')

length111=length(letterlocations3);
length222=1:length111;
length333=length222(letterlocations3);
locations3=randperm(length(length333),ceil(length(length333)*0.1));
locations3=length333(locations3);
pass3=replace(pass3,' ','/');
pass3(locations3)=' ';
writecell({pass3},'rand90 pass3.txt')
```

All statistical comparisons were performed with MATLAB 2019a. Multigroup comparisons were done with Kruskal-Wallis one-way analysis of variance, with the Dunn-Sidak correction.

**Results**

The frequency which each letter of the English alphabet appeared in news articles, novels, plays, and scientific research articles was determined to have some commonality among the categories. The most common letter in all categories was 'E', which varied from a low of 11.91% in news articles to a high of 12.39% in novels (Fig. 1A-D, Table 1). Overall, the six most common letters to appear in all four groups were 'A', 'E', 'I', 'N', 'O', and 'T', of which four were vowels and two were consonants and consisted of ~50% of all letters that appeared. The letter 'Z' was the least common in novels, plays, and scientific articles, ranging from 0.05% in plays to 0.13% in science publications. 'Q', at 0.08%, was the least common letter in news articles. The four least commonly appearing letters in all categories were 'J', 'Q', 'X', and 'Z', all of which were consonants and the sum of which was <1% of all letters used. Scientific publications had the greatest variability in letter usage, as indicated by the large error bars (Fig. 1D). All four categories also exhibited a tri-modal distribution, with peaks at the letters 'E', 'N or O', and 'T'. Collectively, these data demonstrated that distinct types of writing still possess common features, such as most and least used letters and data distribution.

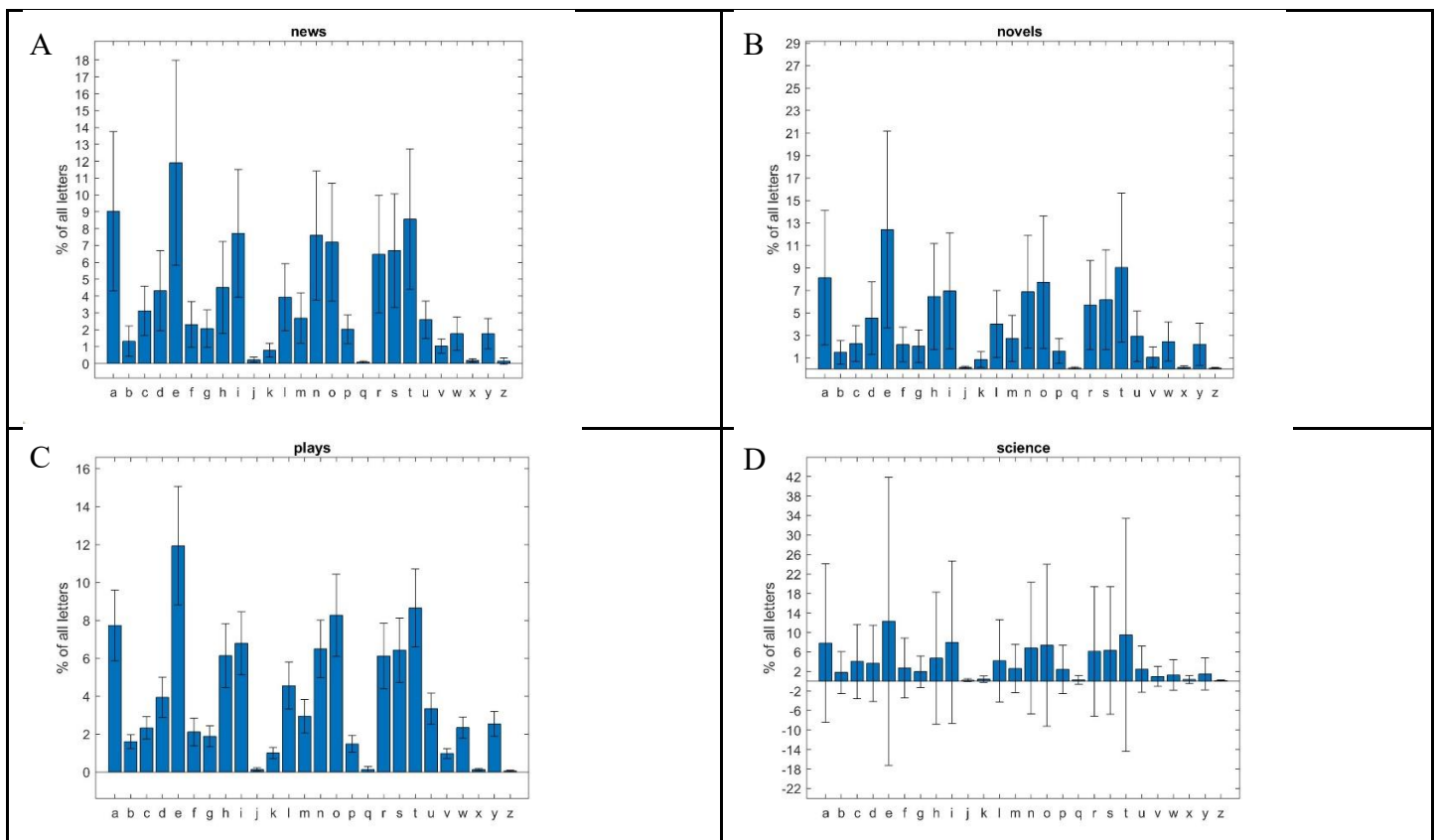

Figure 1: Percentage of each letter, case insensitive, of the English alphabet in different categories of writings. (A) news, (B) novels, (C) plays, and (D) science publications were analyzed for percentage appearance of each letter. n=20 for each category; error bars = standard deviation.

|    | News | Novels | Plays | Science |
|----|------|--------|-------|---------|
| 1  | E (11.91) | E (12.39) | E (11.93) | E (12.32) |
| 2  | A (9.03)  | T (9.03)  | T (8.65)  | T (9.51)  |
| 3  | T (8.56)  | A (8.12)  | O (8.27)  | I (7.96)  |
| 4  | I (7.71)  | O (7.72)  | A (7.73)  | A (7.82)  |
| 5  | N (7.60)  | I (6.94)  | I (6.79)  | O (7.38)  |
| 6  | O (7.20)  | N (6.87)  | N (6.49)  | N (6.82)  |
| 7  | S (6.69)  | H (6.45)  | S (6.43)  | S (6.33)  |
| 8  | R (6.47)  | S (6.16)  | H (6.13)  | R (6.11)  |
| 9  | H (4.51)  | R (5.68)  | R (6.11)  | H (4.73)  |
| 10 | D (4.31)  | D (4.53)  | L (4.55)  | L (4.21)  |
| 11 | L (3.93)  | L (4.00)  | D (3.93)  | C (4.05)  |
| 12 | C (3.13)  | U (2.91)  | U (3.34)  | D (3.68)  |
| 13 | M (2.69)  | M (2.72)  | M (2.94)  | F (2.71)  |
| 14 | U (2.59)  | W (2.43)  | Y (2.54)  | M (2.62)  |
| 15 | F (2.31)  | C (2.26)  | W (2.34)  | U (2.47)  |
| 16 | G (2.07)  | Y (2.20)  | C (2.33)  | P (2.44)  |
| 17 | P (2.03)  | F (2.17)  | F (2.11)  | G (1.96)  |
| 18 | W (1.77)  | G (2.02)  | G (1.88)  | B (1.79)  |
| 19 | Y (1.77)  | P (1.59)  | B (1.60)  | Y (1.49)  |
| 20 | B (1.32)  | B (1.48)  | P (1.48)  | W (1.27)  |
| 21 | V (1.03)  | V (1.04)  | K (1.00)  | V (0.97)  |
| 22 | K (0.78)  | K (0.84)  | V (0.98)  | K (0.40)  |
| 23 | J (0.22)  | X (0.15)  | J (0.14)  | X (0.36)  |
| 24 | X (0.18)  | J (0.13)  | Q (0.13)  | Q (0.28)  |
| 25 | Z (0.14)  | Q (0.10)  | X (0.13)  | J (0.19)  |
| 26 | Q (0.08)  | Z (0.06)  | Z (0.05)  | Z (0.13)  |

Table 1: Percentage of each letter, case insensitive, of the English alphabet in different categories of writings. Each letter with corresponding percentage averages from n=20 selections from each category was ranked from highest to lowest.

In all four groups, ~50% of all letters consisted of the six most frequently used letters; ~70% of all letters consisted of the eight most frequently used letters; and ~90% of all letters consisted of the sixteen most used letters (Fig. 2A-D). Accordingly, 77% of the letters of the alphabet could be removed and still preserve 50% of the information; 69% of the letters of the alphabet could be removed while still preserving 70% of the information; and 38% of the letters of the alphabet could be removed and still preserve 90% of the information. A passage from *I, Pencil: My Family Tree* by Leonard E. Read with the least used 50% of the letters removed and the same passage with a random 50% of letters removed were compared in Fig. 3. Similarly, passages with 30% and 10% of the least used letters removed, along with a random 30% and 10% of letters removed, were compared in Fig. 4 and Fig. 5, respectively. Collectively, these results demonstrated that information is concentrated in a select group of letters in the English language. This observation was consistent across the categories of news, novels, plays, and scientific publications.

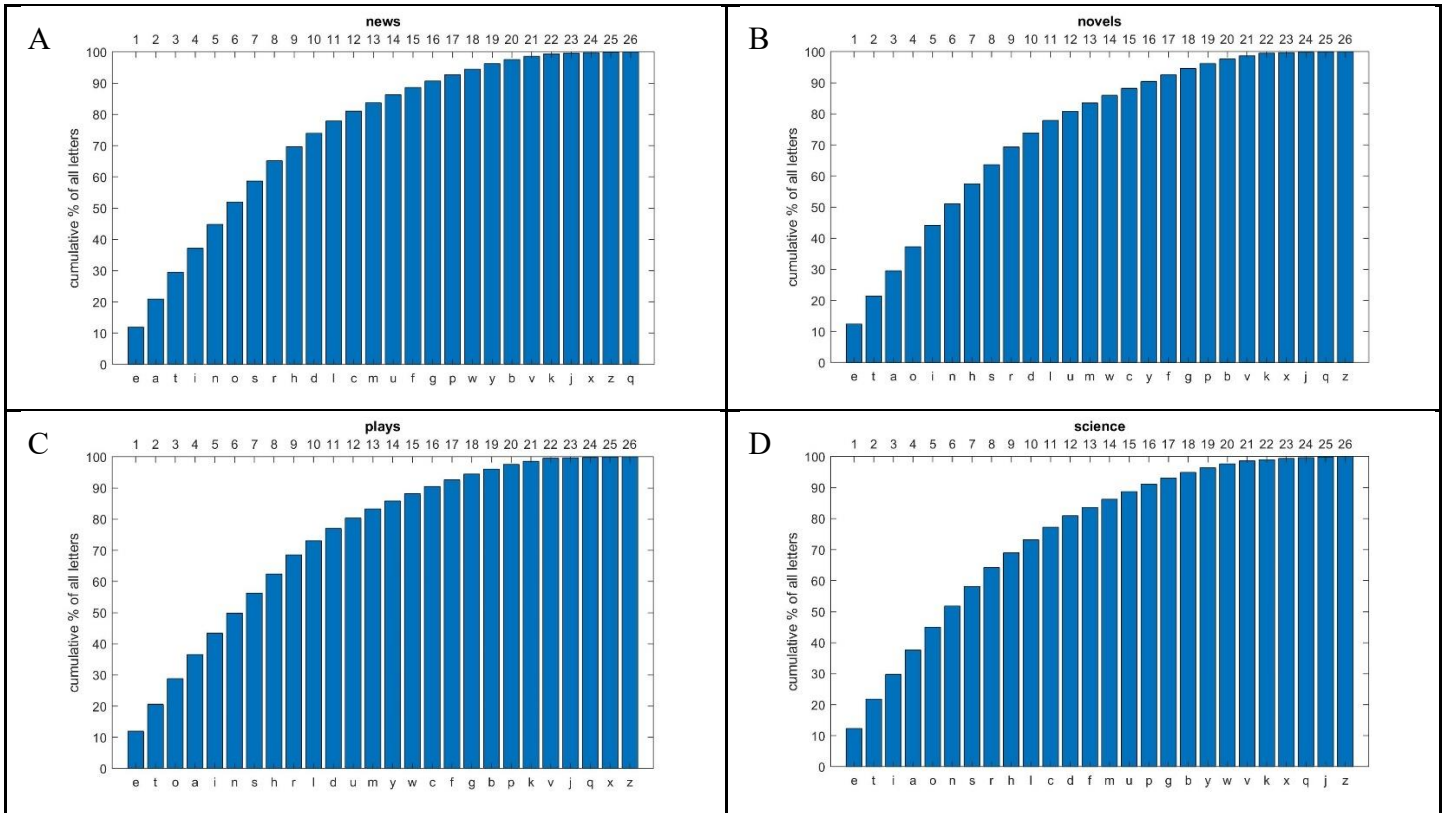

Figure 2: Cumulative percentages of all letters, case insensitive, of the English alphabet in different categories of writings. Cumulative sum of percentages of letters ranked from highest to lowest percentage in n=20 categories of (A) news, (B) novels, (C) plays, and (D) science.

| Original passage |
|---|
| I am a lead pencil—the ordinary wooden pencil familiar to all boys and girls and adults who can read and write. ∗ Writing is both my vocation and my avocation; that's all I do. |
| You may wonder why I should write a genealogy. Well, to begin with, my story is interesting. And, next, I am a mystery—more so than a tree or a sunset or even a flash of lightning. But, sadly, I am taken for granted by those who use me, as if I were a mere incident and without background. This supercilious attitude relegates me to the level of the commonplace. This is a species of the grievous error in which mankind cannot too long persist without peril. For, the wise G. K. Chesterton observed, "We are perishing for want of wonder, not for want of wonders." |
| I, Pencil, simple though I appear to be, merit your wonder and awe, a claim I shall attempt to prove. In fact, if you can understand me—no, that's too much to ask of anyone—if you can become aware of the miraculousness which I symbolize, you can help save the freedom mankind is so unhappily losing. I have a profound lesson to teach. And I can teach this lesson better than can an automobile or an airplane or a mechanical dishwasher because—well, because I am seemingly so simple. |
| Simple? Yet, not a single person on the face of this earth knows how to make me. This sounds fantastic, doesn't it? Especially when it is realized that there are about one and one-half billion of my kind produced in the U.S.A. each year. Pick me up and look me over. What do you see? Not much meets the eye—there's some wood, lacquer, the printed labeling, graphite lead, a bit of metal, and an eraser. |

| Least common 50% erased |
|---|
| I/a /a/ ea / en i —t e/o ina / oo en/ en i / a i ia /to/a / o /an / i /an /a t / o/ an/ ea /an / ite.∗<br>itin /i / ot / / o ation/an / /a o ation;/t at' /a /I/ o.<br> o /a /one / /I/ o / ite/a/ enea o ./e ,/to/ e in/ it ,/ / to /i /inte e tin ./An ,/ne t,/I/a /a' te — o e/ o/t an/a/t ee/o /a/ n et/o /e en/a/ a /o/ i tnin ./ t,/a ,/I/a /ta en/ o / ante/ /t o e/ o/ e/ e,/a /i /I/ e e/a/ e e/in i ent/an / it o t/ a o n ./T i / e i io /attit e/ e e ate / e/ to/t e e /e/o /t e/ o on a e./T i /i /a/ e ie/o /t e/ ie o /e o/ in / i / an in / an not/too/ on / e i t/ it o t/ e i ./ o ,/t e/ i e/ ./ /. e te ton/o e e ,/ " e/a e/ e i in / o / ant/o / on e ,/not/ o / ant/o / on e ." <br>I,/ en i ,/ i e/ to /I/a ea /to/ e,/ e it/ o / one /an /an e,/a/ ai /I/ a atte t/to/ o e./In a t,/i / o / an/ n e tan / e—no,/t at'/ too/ /to/a /o /an one—i / o / an/ e o e/a a e/o /t e/ i a o ne / i /I/ o ie,/ o/ an/ e a e/t e ee o / an in /i / o/ na i / o in ./I/ a e/a/ o o n / e on/to/tea ./An /I/ an/tea /t i / e on/ ette /t an/ an/an/a too ie/o /an/ ai ane/o /a/ e ania /i a e/ e a e— e ,/ ea e/I/a/ ee in / o/ i e. <br>i e?/ et,/not/a / in e/ e on/on/t e/ a e/o /t i /ea t/ no /o /to/ a e/ e./T i o n / anta ti ,/ oe n't/it?/E e ia en/it/i / ea i e /t at/t e e/a e/a o t/one/an /one- a i ion/o / / in / o e /in/t e ..A./ea ea .<br>i / e/ /an / oo /e/o e ./ at/ o/ o / ee?/Not/ / eet/t e /e e—t e e' / o e/ oo ,/ a e ,/t e/ inte /a e in ,/ a ite/ ea ,/a/ it/o / eta ,/an /an/e a e. |
| **50% randomly erased** |
| I/ m/a/ d/p nc l— / di r /oo / n il/ ami iar / o / l/boy /a d/gir s/ / lt / /c /e /an /w it .∗<br>Wr ng/is/bo /my/voc i n/ n / /avocati n;/ a 's/a /I/ o.<br>Y u/m / onde /h /I/ houl r t/ /g ne o y./We ,/to/ w ,/ t y/is/ i ter t ./Ad ,/e ,/I/am/ /m st y— re/s / /te /r/a/ u / or/ v //fa / / ghtn g./But,/ ly,/I/am/ en /o /ra / b /t se/wh / e,/ f /I/e / /mer /in id nt/a d t t/ a kgr nd ./Thi /super ilio / tti u e/r s/me /t e/le el/o /t omm la e./T is/is/a/sp c e / f /th /g i o s/e r / n/ wic / ki c nt/o / on /pe s s /wi h / p ril./For,/t w /G./K./C esterto /o e e ,/ "W ar / e i h / or/ wan / f/ d r,/no /for/wa /of/ de ."<br>I,/Penci ,/s l/ h gh/I/ p t /e,/ rit /your/w r/a /w ,/ cla m/I/ sha t t /o/pro e./I /fa ,/if/y u/ n/u der t nd/me— o,/ ha ' /o /muc /to/ /of /an one— /y c n/ e om/ aw /of/t ac usnes /hi /I/ m oli ,/y/ can/ e p / a e/ h om/ in / / / app y /l i ./I/ /a/ rof u ss /to/t ch./ A d /I/ an/ te ch/this/lesson/ t /than/can/a /a m l/o /a /ir l e/r/a e an c /d s sh /b a se— l,/b cause/I/am/ m n y/o e.<br>Si ?/Y ,/ ot/a/ i gl /pe on/ n/t /fac /o / his/ e r /kn s/ o /t m e/m ./Th o s/fanta , /oe n't/ t?/ Es c l y/ he/ t /s/re l e /th t/t re/ re/ bout/ o / /on -ha / l ion/o /y/k nd/ du d/i /he/U.S.A./ a /y ar.<br>Pick/ e/u / /lo k/ e/o e ./Wha / o/ /se ?/N t/ c /ee s/th / y —th re'/ m /o d,/l q e ,/t e/pri a e i ,/ p /l ad,/ / /o /met ,/ /an/e a er. |

Figure 3: Passage from *I, Pencil: My Family Tree* with the least common 50% of letters and a random 50% of letters removed. "/" represents a space. Letters were case insensitive.

| Original passage |
|---|
| Just as you cannot trace your family tree back very far, so is it impossible for me to name and explain all my antecedents. But I would like to suggest enough of them to impress upon you the richness and complexity of my background.<br>My family tree begins with what in fact is a tree, a cedar of straight grain that grows in Northern California and Oregon. Now contemplate all the saws and trucks and rope and the countless other gear used in harvesting and carting the cedar logs to the railroad siding. Think of all the persons and the numberless skills that went into their fabrication: the mining of ore, the making of steel and its refinement into saws, axes, motors; the growing of hemp and bringing it through all the stages to heavy and strong rope; the logging camps with their beds and mess halls, the cookery and the raising of all the foods. Why, untold thousands of persons had a hand in every cup of coffee the loggers drink!<br>The logs are shipped to a mill in San Leandro, California. Can you imagine the individuals who make flat cars and rails and railroad engines and who construct and install the communication systems incidental thereto? These legions are among my antecedents.<br>Consider the millwork in San Leandro. The cedar logs are cut into small, pencil-length slats less than one-fourth of an inch in thickness. These are kiln dried and then tinted for the same reason women put rouge on their faces. People prefer that I look pretty, not a pallid white. The slats are waxed and kiln dried again. How many skills went into the making of the tint and the kilns, into supplying the heat, the light and power, the belts, motors, and all the other things a mill requires? Sweepers in the mill among my ancestors? Yes, and included are the men who poured the concrete for the dam of a Pacific Gas & Electric Company hydroplant which supplies the mill's power! |

Don't overlook the ancestors present and distant who have a hand in transporting sixty carloads of slats across the nation.

Once in the pencil factory—$4,000,000 in machinery and building, all capital accumulated by thrifty and saving parents of mine—each slat is given eight grooves by a complex machine, after which another machine lays leads in every other slat, applies glue, and places another slat atop—a lead sandwich, so to speak. Seven brothers and I are mechanically carved from this "wood-clinched" sandwich.

My "lead" itself—it contains no lead at all—is complex. The graphite is mined in Ceylon. Consider these miners and those who make their many tools and the makers of the paper sacks in which the graphite is shipped and those who make the string that ties the sacks and those who put them aboard ships and those who make the ships. Even the lighthouse keepers along the way assisted in my birth—and the harbor pilots.

The graphite is mixed with clay from Mississippi in which ammonium hydroxide is used in the refining process. Then wetting agents are added such as sulfonated tallow—animal fats chemically reacted with sulfuric acid. After passing through numerous machines, the mixture finally appears as endless extrusions—as from a sausage grinder-cut to size, dried, and baked for several hours at 1,850 degrees Fahrenheit. To increase their strength and smoothness the leads are then treated with a hot mixture which includes candelilla wax from Mexico, paraffin wax, and hydrogenated natural fats.

My cedar receives six coats of lacquer. Do you know all the ingredients of lacquer? Who would think that the growers of castor beans and the refiners of castor oil are a part of it? They are. Why, even the processes by which the lacquer is made a beautiful yellow involve the skills of more persons than one can enumerate!

Observe the labeling. That's a film formed by applying heat to carbon black mixed with resins. How do you make resins and what, pray, is carbon black?

My bit of metal—the ferrule—is brass. Think of all the persons who mine zinc and copper and those who have the skills to make shiny sheet brass from these products of nature. Those black rings on my ferrule are black nickel. What is black nickel and how is it applied? The complete story of why the center of my ferrule has no black nickel on it would take pages to explain.

Then there's my crowning glory, inelegantly referred to in the trade as "the plug," the part man uses to erase the errors he makes with me. An ingredient called "factice" is what does the erasing. It is a rubber-like product made by reacting rape-seed oil from the Dutch East Indies with sulfur chloride. Rubber, contrary to the common notion, is only for binding purposes. Then, too, there are numerous vulcanizing and accelerating agents. The pumice comes from Italy; and the pigment which gives "the plug" its color is cadmium sulfide.

**Least common 30% erased**

 st/as/ o / annot/tra e/ o r/ a i /tree/ a / er/ ar,/so/is/it/i ossi e/ or/ e/to/na e/an /e  ain/a / /ante e ents./ t/I/ o   / i e/to/s est/eno  h/o /the /to/i  ress/ on/ o /the/ri hness/an / o  e it /o / / a  ro n .

 / a i /tree/ e ins/ ith/ hat/in/ a t/is/a/tree,/a/ e ar/o /strai ht/ rain/that/ ro s/in/Northern/ a i ornia/an /Ore on./No / onte ate/a /the/sa s/an /tr  s/an /ro e/an /the/ o nt ess/other/ ear/ se /in/har estin /an / artin /the/ e ar/ o s/to/the/rai roa /si in ./Thin o /a /the/ ersons/an /the/n  er ess/s i s/that/ ent/into/their/ a ri ation:/the/ inin /o /ore,/the/ a in /o /st ee /an /its/re ine ent/into/sa s,/a es,/ otors;/the/ ro in /o /he  an / rin in /it/thro  h/a /the/sta es/to/hea  an /stron /ro e;/the/ o in / a s/ ith/their/ e s/an / ess/ha s,/the/ oo er/an /the/raisin /o /a /the/ oo s./ h ,/ nto / tho san s/o / ersons/ha /a/han /in/e er /o / o  ee/the/ o ers/ rin !

The/ o s/are/shi e /to/a /i in/San/ ean ro,/ a i ornia./ an/ o /i a ine/the/in i i a s/ ho/ a e/ at/ ars/an /rai s/an /rai roa /en ines/an / ho / onstr t/an /insta  /the/ o ni ation/s ste s/in i enta /thereto?/These/ e ions/are/a on / /ante e ents.

 onsi er/the/i  or /in/San/ ean ro./The/ e ar/ o s/are/  t/into/s a ,/ en i - en th/s ats/ ess/than/one- o rth/o /an/in h/in/thi ness./These/are/ i n/ rie /an /then/tinte / or/the/sa e/reason/ o en/  t/ro e/on/their/ a es./ eo  e/ re er/that/I/ oo / rett ,/not/a/ a i / hite./The/s ats/are/ a e /an / i n/ rie /a ain./Ho / an /s i s/ ent/into/the/ a in /o /the/tint/an /the/ i ns,/into/s  in /the/heat,/the/ i ht/an / o er,/the/  ets,/  otors,/an /a /the/other/thin s/a/ i  re ires?/S ee ers/in/the/ i /a on / /an estors?/ es,/an /in  e /are/the/ en/ ho/ o re/the/ on rete/ or/the/ a /o /a/ a i i / as///E e tri / o  an /h ro  ant/ hi h/s  ies/the/ i  's/ o er!

 on't/o er oo /the/an estors/ resent/an / istant/ ho/ha e/a /han /in/trans ortin /si t / ar oa s/o /s ats/a ross/the/nation.

On e/in/the/ en i / a tor —$4,000,000/in/ a hiner /an / i in ,/a / a ita /a   ate / /thri t /an /sa in / arents/o / ine—ea h/s at/is/ i en/ei ht/ roo es/  /a/ o   e / a hine,/a ter/ hi h/another/ a hine/ a s/ ea s/in/e er /other/s at,/a  ies/  e,/an / a es/another/s at/ato —a/ ea /san i h,/so/to/s ea ./Se en/ rothers/an /I/are/ e hani a   / ar e / ro /this/" oo - in he "/san i h.

  /" ea "/itse  —it/ ontains/no/ ea /at/a  —is/ o  e ./The/ ra hite/is/ ine /in/ e on./ onsi er/these/ iners/an /those/ ho/ a e/their/ an /too s/an /the/ a ers/o /the/ a er/sa s/in/ hi h/the/ ra hite/is/shi  e /an /those/ ho/ a e/the/strin /that/ties/the/sa s/an /those/ ho/ t/the /a oar /shi s/an /those/ ho/ a e/the/shi s./E en/the/ i htho se/ ee ers/a on /the/ a /assiste /in/ / irth—an /the/har or/ i ots.

The/ ra hite/is/ i e / ith/ a /ro /ississi i/in/ hi h/a oni /h ro i e/is/ se /in/the/re inin / ro ess./Then/ ettin /a ents/are/a e /s h/as/s onate /ta o —ani a /ats/ he i a /rea te / ith/s ri i a i./A ter/ assin /thro h/n ero s/ a hines,/the/ i t re/ ina /a ears/as/en ess/e tr sions—as/ ro /a/sa sa e / rin er- t/to/si e/ rie ,/an / a e / or/se era /ho rs/at/1,850/ e rees/ ahrenheit./To/in rease/their/stren th/an /s oothness/the/ ea s/are/then/treate / ith/a/hot/ i t re/ hi h/in es/ an e i a/ a / ro / e i o,/ ara in /a ,/an /h ro enate /nat ra / ats.

/ e ar/re ei ies/si /oats/o /a er./ o/ o /no /a /the/in re ients/o / a er?/ ho/ o /thin /that/the/ ro ers/o / astor/ eans/an /the/re iners/o / astor/oi /are/a / art/o /it?/The /are./ h ,/e en/the/ ro esses/ / hi h/the/ a er/is/ a e/ ea ti / e o /in o e/the/s i s/o / ore/ ersons/than/one/ an/en erate!

O ser e/the/ a e in./That's/a / i / ore / / a in /heat/to/ ar on/ a / i e/ ith/resins./Ho / o / a e/resins/an / hat,/ ra ,/is/ ar on/ a ?

/ it/o / eta —the/ err e—is/ rass./Thin /o /a /the/ ersons/ ho/ ine in /an / o er/an /those/ ho/ha e/the/s i s/to/ a e/shin /sheet/ rass/ ro /these/ ro ts/o /nat re./Those/ a /rin s/on/ err e/are/ a /ni e./ hat/is/a /a ni e an /ho /is/it/a ie ?/The/ o ete/stor /o / h /the/ enter/o / / err e/has/no / a /ni e/on/it/ o / ta e/ a es/to/e ain.

Then/there's/ ro nin / or ,/ine e ant /re erre /to/in/the/tra e/as/"the / ,"/the/ art/ an ses/to/erase/the/errors/he/ a es/ ith/ e./An/in re ient/ a e /"a ti e"/is/ hat/ oes/the/erasin./It/is/a/r er- i e ro t/ a e / rea tin /ra e-see /oi / ro /the/ t h/East/In ies/ ith/s r h ori e./R er,/ ontrar /to/the/ o on/notion,/is/on / or/ in in / r oses./Then,/too,/there/are/n ero s/ ani in /an /a e eratin /a ents./The / i e/ o es/ ro /Ita ;/an /the/ i ent/ hi h/ i es/"the / "/its/ o or/is/ a i /s i e.

30% randomly erased

Just/ s/ /cannot/ ce/y ur/family/tree/ba k/v y ar,/so/is/i /i possi le/f r/me/ nam /an /ex la /a l/my/ant e ent ./B /I/ oul /lik /t / ugge oug / f/ /to/i p e /upon/y u/th /ri hne s/a / omp ex ty/of/my/bac round.
M /f mily ee e ns/with/w at/ n/fact/i /a ree,/ / e ar /stra ght/grain/that/gr w /in/Nor hern/Cali ornia/ /Oregon./No w/ n e p te/ ll/the/ a /and/ ucks/ nd/ /an he/cou tless/other/ge use / n/ rvestin / d/carti g/the/ e r ogs/t he r il o d/s din ./Thi k/of/ ll/the/p rso s/an t /numberl ss/s ills/th t/ t/into/their/ ric ion:/ he/ ining/of/ ore,/the/mak g/of/steel/and/its/refin ent/i o/saw ,/ es,/moto ;/he wing/ hem /an /b ingi g/i through/all/the/sta es/to/ he vy/and/s on /rop ;/he og ng/cam s/with/ hei /b ds/and/ ss l ,/the/co ke y/an /th /rai in / f/ ll/ he/fo ds./Why,/un ld/ ousa d /of/p rsons/ ad/a/hand/in/eve y/c /of/c ff e/ the/ oggers/d in !
T e/lo s/a e hip ed/to/a/mill/in/Sa /Lea dro,/Califo a./C n/y u/i g n/th /i divi al/wh /make/ l ca s/ d/ra ls/a d/r lro d/e g e/ an /w o/con truct/ nd/ t ll/the/commu i at on/sys ems/ nc den al/thereto?/T ese/le io a /amo g/my/ n ce ts.
Co d /the/mi lw rk/ n/San/L a dro./Th ce r/lo s/are/cut/it /all,/pen i -length/s at es/tha o e-f ur h/ a /nc i /thick es ./These/a il drie /nd/th n te/fo /he ame/easo me/p /roug n/thei aces./Peopl pr r/ta /I/loo e ,/ t/a/palli / hi e./Th las/are/wa ed/and/ iln/dri d/ g ./Ho ma /skil s/wen n o/ m ki g/of/ e t nt/ nd/the/kil s,/in o/supp y g/ he at,/the/l ght/and/po e ,/the/bel s,/m ors,/a d/all/th her ing /mi /req r ?/Sw epers/in/the/mi l/amon / y/ nce to s?/Ye ,/ d/in l ded/ re/t m n/who/ ured/ he co ete/f r/the/da /of/a/P f c/Gas/&/Electric/Com y/ yd p a t/ ich/s p li s/ he/mil '/ ower!
Do 't/o e o k/the/ esto s/pr sent/ n /dista t/ ve/a/ha /in/t an po ti s xty/ arloa /of/ las/ r ss/te/ ati n.
Onc / n/t p nc /fact ry—$4,000,000/ /machinery/and/bui ding,/a l/capit l/acc m lat d/b /thrif /a s ng/parents/of/mine—e c /slat s/given/eig t/ r oves/ y/a/co pl x/ ac ie ,/ fter/wh h/ othe /ma hi e/lays/l a /in e y/othe /sla ,/a lie /glue,/and/pl ce /another/slat/a op— /lead/sandwich,/ /o/s e k./Seven/b ot er /and/I/a e/mech ni ally/carved/f om/ his/"wood-c in ed"/sandwi h.
My/" "/ tsel —it/ ontain /o/ l ad/a /a l— /c mplex./Th r e/is/min d/i /Ceyl ./Cons de t ese/miners/and/t ose/w /ma /thei /many/t ols/ d/t e/ akers/ f/the/p p /sac s/ n/ h h/the/gra h t/is/ h pped/a t e/w /ma th / ri g t/at/ ies/th /sac s/ nd/those/who/ ut/th / o d/ships/ nd/t o e/who/m k /th /ships./Even/t e/l ght s / eepes/ l t e/way/ ssisted/i m / r t —a d/th /ha bor/pilots.
T e/graph te/i ixe /w h/c ay/fro /Mis issi i/i /hich/am o iu /hydr x e/is/ se /n t e/refini / roces ./Then/w tt ng/age s/are/added/su h/as/sulfo te /tallo —a i al/ / emic l y/ eacted/ t/sulfu ic/a id./A ter/ s ng/through/nu ero s/ma hin s,/th mixtu e/fina ly/a es/ s/s e le s/ x usio s—as f o /a/ ausag r nde -cut/to/si e,/dri d,/and/baked/ or/ ever / ou s/at/1,850/d gree /Fa nhei ./T ncrease/t ir/ trength/a /moothn /he/le ds/are/ hen/t a ted/w t /hot/m xture/w ich/includes/ andel la / f om/M ico,/ raff n/w x,/a d/hyd og n ted/na ural/fa s.
My/ced r/r c es/s /co t /of/lacquer./D yo /k ow/all/th redien /of/la qu ?/Who/ ould/t ink/that/th g o ers/of/c st / ans/an t / r finers/o /cas r/oil/ e/a art/of/ t?/They/are./Wh ,/eve /the/ ocesse /by/ hich/t e/ acquer/i m de/a/be u iful y low/in lve/t e/sk ls/of/mor /pe s ns/ han/o e c n/e um r te!

| |
|---|
| Ob  r e/th / abe in ./That's/ /film/formed/ /app y g/ eat/to/c rbo /bl c /mixe / ith/resins./How/ o/ o /m k / es ns/and/ ha ,/pray,/is/car n/bl ck? |
| My/b t/ /metal—the/ er ule—is/br s ./Th n /o / l / he/ er s/ h /mine/ nc/and/copper/ d/th  /who/ha /th /s i ls/to/make/sh n / heet/bras /from/t es /p oducts/o /na re./T ose/ ack/ i on/my/ e le/ re/ ac /n cke ./W a /is/ lack/nickel/an / ow/is/it/appl ed?/The/co pl  / t ry/of/wh  e/c nt r/of/ /f r ule/h s/ /black/ ic el/on/it/w u d/  e/ g s/t /expl n. |
| Th n/ e 's/m / rown ng/glor ,/i e ega tl r ferred/to/in t e/ r de/as/" he/p  ,"/t e/part/m n/ s/ o/er e/the/error /he/ mak s/wi h/ e./A /ing edient/ lle /" cti "/is/wha /doe / h /era in ./It/i /a/rub  -l ke/pr du t/mad / y/rea ting/ra e- eed/oil/f m/th /Dut / East/Indies/wi h/sul  / hlori e./Rub er,/co trary/ o/ e/co mon/no   ,/is/only/fo /bin ing/p rpo e ./Then,/too,/th r /a e/ me o / ulcan zing/and/ ccel  ing/agen s./T /pumice/ omes/ rom/I a y;/and/t e/pi en /wh ch/gi /" e/ ug"/its/ or/is/ca mium/ lf de. |

Figure 4: Passage from *I, Pencil: My Family Tree* with the least common 30% of letters and a random 30% of letters removed. "/" represents a space. Letters were case insensitive.

| **Original passage** |
|---|
| Does anyone wish to challenge my earlier assertion that no single person on the face of this earth knows how to make me? |
| Actually, millions of human beings have had a hand in my creation, no one of whom even knows more than a very few of the others. Now, you may say that I go too far in relating the picker of a coffee berry in far off Brazil and food growers elsewhere to my creation; that this is an extreme position. I shall stand by my claim. There isn't a single person in all these millions, including the president of the pencil company, who contributes more than a tiny, infinitesimal bit of know-how. From the standpoint of know-how the only difference between the miner of graphite in Ceylon and the logger in Oregon is in the type of know-how. Neither the miner nor the logger can be dispensed with, any more than can the chemist at the factory or the worker in the oil field—paraffin being a by-product of petroleum. |
| Here is an astounding fact: Neither the worker in the oil field nor the chemist nor the digger of graphite or clay nor any who mans or makes the ships or trains or trucks nor the one who runs the machine that does the knurling on my bit of metal nor the president of the company performs his singular task because he wants me. Each one wants me less, perhaps, than does a child in the first grade. Indeed, there are some among this vast multitude who never saw a pencil nor would they know how to use one. Their motivation is other than me. Perhaps it is something like this: Each of these millions sees that he can thus exchange his tiny know-how for the goods and services he needs or wants. I may or may not be among these items. |
| **Least common 10% erased** |
| Does/anyone/wish/to/challen e/my/earlier/assertion/that/no/sin le/ erson/on/the/ ace/o /this/earth/ nows/how/to/ma e/me? |
| Actually,/millions/o /human/ ein s/ha e/had/a/hand/in/my/creation,/no/one/o /whom/e en/ nows/more/than/a/ ery/ ew/o /the/others./Now,/you/may/say/that/I/ o/too/ ar/in/relatin /the/ ic er/o /a/co ee/ erry/in/ ar/o / ra il/and/ ood/ rowers/elsewhere/to/my/creation;/that/this/is/an/e treme/ osition./I/shall/stand/ y/my/claim./There/isn't/a/sin le/ erson/in/all/these/millions,/includin /the/ resident/o /the/ encil/com any,/who/contri utes/more/than/a/tiny,/in initesimal/ it/o / now-how./ rom/the/stand oint/o / now-how/the/only/di erence/ etween/the/miner/o / ra hite/in/Ceylon/and/the/lo er/in/Ore on/is/in/the/ty e/o / now-how./Neither/the/miner/nor/the/lo er/can/ e/dis ensed/with,/any/more/than/can/the/chemist/at/the/ actory/or/the/wor er/in/the/oil/ ield— ara in/ ein /a/ y- roduct/o / etroleum. |
| Here/is/an/astoundin / act:/Neither/the/wor er/in/the/oil/ ield/nor/the/chemist/nor/the/di er/o / ra hite/or/clay/nor/any/who/mans/or/ma es/the/shi s/or/trains/or/truc s/nor/the/one/who/runs/the/machine/that/does/the/ nurlin /on/my/ it/o /metal/nor/the/ resident/o /the/com any/ er orms/his/sin ular/tas / ecause/he/wants/me./Each/one/wants/me/less,/ erha s,/than/does/a/child/in/the/ irst/ rade./Indeed,/there/are/some/amon /this/ ast/multitude/who/ne er/saw/a/ encil/nor/would/they/ now/how/to/use/one./Their/ moti ation/is/other/than/me./ erha s/it/is/somethin /li e/this:/Each/o /these/millions/sees/that/he/can/thus/e chan e/his/tiny/ now-how/ or/the/ oods/and/ser ices/he/needs/or/wants./I/may/or/may/not/ e/amon /these/items. |
| **10% randomly erased** |
| Does/a yon /w sh/to/ch llenge/my/earlier/asse tion/that/no/single/person/on/the/face/of/this/e th/knows/h w/to/make/m ? |

> Actuall ,/mil ion /of/human/beings/have/   /a/ and/in/my/creation,/no/one/of/ hom/even/knows/m re/than/a/very/ ew/of/the/others./Now,/you/may/say/that/I/g /too/far/in/rela ing/the/pick r/of/a/coffee/berr /in/far/off/Brazil/and/food/grower /elsewhe /to/my/ r ation;/that/this/is/an/extr me/posit on./I/sha l/stand/by/my/cla m./Ther /isn't/a/sing e/pe s n/i /all/th se/millions,/inc uding/the/president/of/t e/pencil/company,/who/contributes/mo e/th n/a/tiny,/infinitesi l/bi /of/know-h w./Fr m/the/standp int/of/know-how/the/on y/difference/betw en/the/min r/o /graphite/in/Ceylo /and/the/logger/in/Oregon/is/in/the/type/of/know-how./Neither/the/miner/ r/the/ ogge /ca /be/dispensed/with,/any/ ore/than/ an/the/chemis /at/the/ a t r /o /the/wo ker/in/the/ i /field—para fin/b ing/a/by-product/of/petroleum.
> Her /is/an/astounding/fac :/Neith r/the/w r er/in/t e/oil/field/nor/th /c e st/nor/the/digger/of/graphite/or/clay/ or/any/who/ ans/or/makes/the/ hips/or/tr i /or/trucks/no /the/one/who/runs/ h /mac ine/t at/ oes/the/ nurling/on/ y/b t/of/metal/nor/the/president/of/the/company/p rforms/his/singula /ta /b cause/he/wa ts/me./Ea h/one/want /me/less,/perhaps,/than/does/ /ch ld/i /the/first/ rade./I deed,/there/are/some/among/ his/vast/multitude/ ho/never/saw/ /pencil/nor/wou d/they/know/h w/to/use/on ./Thei /motivation/is/ot er/than/m ./Perhaps/it/is/som thing/like/ his:/Each/of/these/milli ns/sees/that/ e/can/thus/exchang /his/tiny/know-how/for/the/ od /and/servic s/he/needs/or/wants./I/may/or/m y/not/be/among/t ese/item .

Figure 5: Passage from *I, Pencil: My Family Tree* with the least common 10% of letters and a random 10% of letters removed. "/" represents a space. Letters were case insensitive.

We next applied our analysis of letter frequency to predicting writing category. The percentage frequency which each letter of the English alphabet appeared in test news articles was compared with standards developed from evaluating the letter frequencies of known news articles, novels, plays, and scientific articles. The distance between the test news articles and the news standard was the least, indicating greatest similarity of the four categories. However, no significant difference was measured among the distances with any of the standards (Fig. 6A, Table 2). The distance between test novels and the novel standard was the least and significantly lower than that with the news or science standards (Fig. 6B, Table 3). The test plays exhibited the least distance with the play standard (Fig. 6C, Table 4). The test science articles exhibited the least distance with the science standard and was significantly lower than distance with the novel or play standards (Fig. 6D, Table 5). Collectively, this analysis showed that news articles, novels, plays, and science articles possess deterministic percentage letter frequencies. Furthermore, this data could be synthesized into a predictive algorithm for determining writing categories.

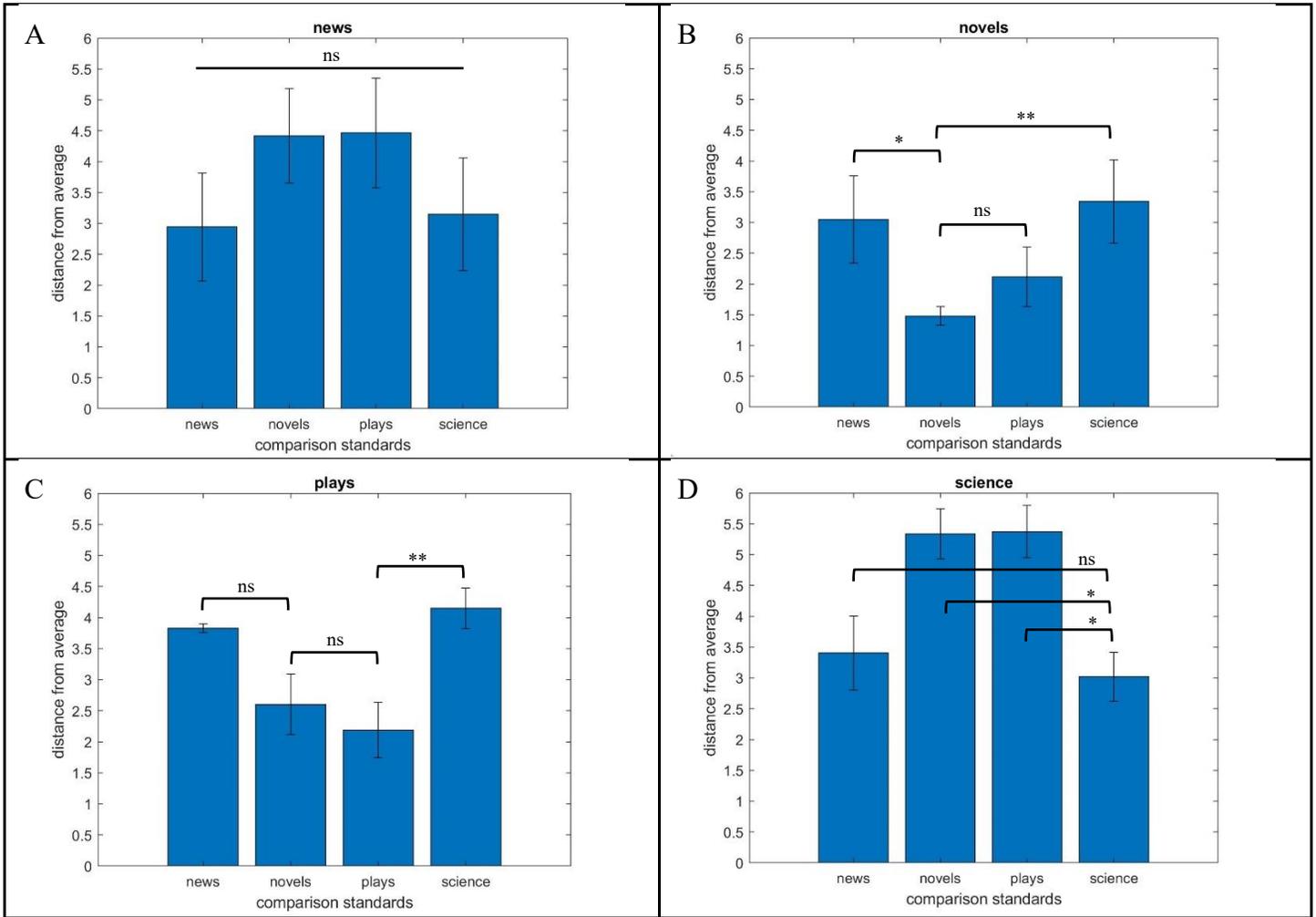

Figure 6: Using the percentage frequency of each letter, case insensitive, of the English alphabet as a predictive metric for categorizing (A) news, (B) novels, (C) plays, and (D) science publications. Distance between the percentage frequency of letters in (A) test news articles, (B) test novels, (C) test plays, (D) test science articles and the average percentage frequency of letters in n=20 writings from different comparison standards. A higher distance is indicative of greater dissimilarity, while lower distance being indicative of greater similarity, between the test writing and the comparison standard. n=5 for each test category; error bars = standard deviation. Comparisons performed using the Kruskal-Wallis test with the Dunn-Sidak correction. ns = $p>0.05$; * = $p<0.05$; ** = $p<0.01$.

|                  | News test 1 | News test 2 | News test 3 | News test 4 | News test 5 |
|------------------|-------------|-------------|-------------|-------------|-------------|
| News standard    | 2.18        | 3.82        | 3.91        | 2.69        | 2.10        |
| Novel standard   | 3.67        | 5.50        | 4.84        | 4.30        | 3.78        |
| Play standard    | 3.71        | 5.97        | 4.46        | 4.21        | 3.97        |
| Science standard | 1.78        | 4.06        | 3.90        | 3.05        | 2.95        |

Table 2: Distance between percentage letter frequencies of test news articles and average percentage letter frequencies of n=20 comparison standards from different categories of writings. Letters were case insensitive.

|  | Novel test 1 | Novel test 2 | Novel test 3 | Novel test 4 | Novel test 5 |
| --- | --- | --- | --- | --- | --- |
| News standard | 2.24 | 3.99 | 2.77 | 2.69 | 3.55 |
| Novel standard | 1.49 | 1.69 | 1.40 | 1.53 | 1.29 |
| Play standard | 2.16 | 2.55 | 1.29 | 2.28 | 2.32 |
| Science standard | 2.73 | 4.30 | 3.13 | 2.78 | 3.77 |

Table 3: Distance between percentage letter frequencies of test novels and average percentage letter frequencies of n=20 comparison standards from different categories of writings. Letters were case insensitive.

|  | Play test 1 | Play test 2 | Play test 3 | Play test 4 | Play test 5 |
| --- | --- | --- | --- | --- | --- |
| News standard | 3.88 | 3.91 | 3.82 | 3.74 | 3.80 |
| Novel standard | 3.12 | 2.63 | 2.17 | 2.05 | 3.04 |
| Play standard | 2.42 | 1.87 | 1.81 | 2.86 | 1.98 |
| Science standard | 3.92 | 4.64 | 3.91 | 4.32 | 3.95 |

Table 4: Distance between percentage letter frequencies of test plays and average percentage letter frequencies of n=20 comparison standards from different categories of writings. Letters were case insensitive.

|  | Science test 1 | Science test 2 | Science test 3 | Science test 4 | Science test 5 |
| --- | --- | --- | --- | --- | --- |
| News standard | 3.75 | 4.07 | 2.50 | 3.50 | 3.21 |
| Novel standard | 5.65 | 5.82 | 4.79 | 5.26 | 5.16 |
| Play standard | 5.82 | 5.74 | 4.83 | 5.41 | 5.07 |
| Science standard | 2.97 | 3.54 | 2.74 | 2.56 | 3.28 |

Table 5: Distance between percentage letter frequencies of test science articles and average percentage letter frequencies of n=20 comparison standards from different categories of writings. Letters were case insensitive.

**Discussion**

A vast category of writings exists in the English language. However, the commonality of sharing a single English alphabet reveals patterns that can be used to quantify and evaluate their underlying characteristics. The frequency with which each letter of the alphabet appears gives rise to a tri-modal distribution that is shared among news articles, novels, plays, and science publications. Perhaps magnitude of deviation from this distribution could be indicative of external influence not typically encountered in regular writing. This distribution can also be rearranged to reveal the most and least commonly used letters. Consequentially, one observes that ~23% of the letters of the alphabet is used to populate ~50% of letters used in writings; ~62% of letters is used to populate ~90% of used letters. Such a phenomenon would naturally have an application to the transmission of information under circumstances with limited resources.

The average percentage frequency with which each letter appears in different categories of writing also gives rise to a potential mechanism for writing categorization. The metric of distance, d, which we defined as magnitude of separation from the average percentage frequency calculated in a standard (mathematically defined in methods) was successful in distinguishing news articles, novels, plays, and scientific articles. More robust standards could be developed by increasing the number of predetermined writings of which the average percentage frequency of letters is calculated. The ability to automatically categorize writings could find application in curating large informational databases. A simple algorithm, whether based on our distance metric or another set of metrics, to recognize article type would find use in online scientific journals or other publication platforms that handle heavy user traffic and are unable to allocate sufficient manpower for timely manual evaluation of all submitted material.

We have shown that writings from different categories possess unique determinants yet also share common characteristics that can be readily extracted. The results from this manuscript can be used to further the fields of linguistics, information communication, and large data processing.

# References

[1]  Anon Letter Frequencies in the English Language

[2]  Smith L D 1955 *Cryptography The Science of Secret Writing* (Dover Publications)

[3]  Pratt F 1942 *Secret and Urgent: The Story of Codes and Ciphers* (Blue Ribbon Books)

[4]  Ohaver M E 1933 *Cryptogram Solving* (Stoneman Press)

[5]  Griffith R T 1949 The minimotion typewriter keyboard *J Franklin Inst* **248** 399–436

[6]  Bourne C P and Ford D F 1961 A study of the statistics of letters in English words *Information and Control* **4** 48–67